\documentclass[10pt,conference,compsocconf]{IEEEtran}
\IEEEoverridecommandlockouts
\usepackage{times}

\usepackage{caption}
\usepackage{cite}

\ifCLASSINFOpdf
  \usepackage[pdftex]{graphicx}
  \DeclareGraphicsExtensions{.pdf,.jpeg,.png}
\else
  \usepackage[dvips]{graphicx}
  \DeclareGraphicsExtensions{.eps}
\fi
\usepackage{amsmath}
\usepackage{subfigure}

\usepackage{url}

\usepackage{textcomp}
\usepackage{gensymb}

\begin{document}
\pagenumbering{gobble}
\bstctlcite{IEEEexample:BSTcontrol}
%
\title{\textbf{\Large Geographic Centroid Routing for Vehicular Networks}}

\author{\IEEEauthorblockN{Justin P. Rohrer}
\IEEEauthorblockA{Naval Postgraduate School, Monterey, CA 93943, USA\\
email: jprohrer@nps.edu}
\thanks{This paper is authored by employees of the United States Government and
is in the public domain. Non-exclusive copying or redistribution is allowed,
provided that the article citation is given and the authors and agency are clearly
identified as its source.  Approved for public release: distribution unlimited.}
}


%


\maketitle

\begin{abstract}
A number of geolocation-based Delay Tolerant Networking (DTN) routing 
protocols have been shown to perform well in selected simulation and 
mobility scenarios.  However, the suitability of these mechanisms for 
vehicular networks utilizing widely-available inexpensive Global Positioning System (GPS) hardware has not been evaluated.
We propose a novel geolocation-based routing primitive (Centroid Routing) that is resilient
to the measurement errors commonly present in low-cost GPS devices.
Using this notion of Centroids, we construct two novel routing protocols and evaluate their
performance with respect to positional errors as well as traditional DTN routing metrics.
We show that they outperform existing approaches by a significant margin.
\end{abstract}


%
\IEEEpeerreviewmaketitle

\section{Introduction}
Research published by the Delay- and Disruption-Tolerant Networking (DTN) community over the last decade shows significant benefits to incorporating geolocation information into routing algorithms.  This is unsurprising, given that DTN routing protocols are required to make local forwarding decisions, without the benefit of consistent global routing information.  

Much of this work is evaluated only in simulation and emulation environments (and we include our own prior work in making this generalization~\cite{rohrer:2008:CLA}), in which the positional measurements are assumed to be highly accurate.  In practice, vehicular communication modules are often (and perhaps increasingly so) constructed from very inexpensive hardware without high-quality antennas or complex GPS chipsets, and expected to function in urban canyons or other environments with partially obstructed GPS signals.

Under such conditions, the advertised $\pm$20~m civilian GPS accuracy bounds quickly decay to nearly 200 m, with the more eccentric error typically occurring orthogonally to the direction of travel, without resembling a normal error distribution~\cite{benmoshe:2011:IAO}.  The implication for the consumer of such position data is that the location delta between updates due to error may be an order of magnitude larger than the actual distance travelled in the same time.  Simply taking additional samples cannot resolve this error due to the high correlation between consecutive GPS location readings.  This explains the all-too-common scenario of ``my GPS thinks I'm driving in a field/lake/building/offramp/etc''.  Commercial GPS-based mapping devices are relatively successful at hiding such inaccuracies by taking hints from the map database and making sophisticated assumptions (learned through decades of development on this single application), such as smoothed travel trajectories and snapping the position to nearby roads.  However, when these assumptions are wrong, even greater errors may be introduced so we must find other mechanisms for mitigating the underlying errors in positional data.  Please note that we don't mean to imply that advances in technology won't decrease these errors; technology trickle-down, availability of Global Navigation Satellite System (GLONASS), and planned improvements in future GPS satellites will all have that effect in the coming decades, however the current general assumption of zero error will continue to be unwarranted for the foreseeable future.

Our contributions in this work include a novel routing primitive and two novel routing protocols based on this primitive.  We also perform an analysis of the effects of errors in positional data on our two protocols and an existing protocol.  Lastly, we contribute an oracle router for the ONE simulator.  Code for all routers is made available via the Tactical Networked Communication Architecture Design lab website~\cite{tancad:www}.

The structure of this paper is as follows:  Section~\ref{sec:rw} discusses the prior work in DTN routing protocols that we build upon in this work.  Section~\ref{sec:cbr} presents our new routing primitive, and two DTN routing protocols based on that primitive.  Section~\ref{sec:sa} evaluates the protocols via the ONE Simulator.  Section~\ref{sec:c} concludes.

\section{Related Work}\label{sec:rw}
Our routing primitive design takes inspiration from a number of location-based routing protocols developed for mobile ad-hoc networks,
including Vector, APRAM, DREAM, SIFT, and GRID~\cite{kang:2008:VRF,liao:2001:GAF,fuente:2007:APC,galluccio:2007:AMR,mauve:2001:ASO,yuksel:2006:AIF}.  
Vector maximizes message spreading by preferentially transferring messages to neighbors traveling in a direction orthogonal to the node's own direction of travel.  It calculates the trajectory vectors from repeated GPS samples.
APRAM~\cite{iordanakis:2006:ARP} utilizes GPS coordinates to discover the geographically shortest path to the destination, while DREAM uses the cached 
node locations to make local forwarding decisions that forward packets in the direction of the destination.  Similarly, AeroRP~\cite{jabbar:2009:AAG}\hspace{1sp}\cite{rohrer:2011:HDC}\hspace{1sp}\cite{sterbenz:2018:DTA}
uses both the coordinates and velocity of neighbors to locally determine the best next hop.  LAR~\cite{ko:2000:LAR} uses location information to bound the area of the route discovery phase, 
thus reducing overhead.  Beaconless geographic routing~\cite{sanchez:2009:BGR} exploits the broadcast nature of wireless channels to overhead the location of neighboring nodes, 
and use this information to discover the best route.  Other protocols such as IGF~\cite{blum:2003:IAS}, BOSS~\cite{sanchez:2007:BBO}, and BLR~\cite{heissenbuttel:2004:BBR} have been proposed that vary in the algorithm used to select the forwarding node.  

We presented a 2-page poster paper describing the issue of ignoring GPS errors when simulating geographic grouting protocols~\cite{rohrer:2017:EOG}, however it did not include our novel protocols or the simulation analysis described in this work.

\section{Centroid-Based Routing}\label{sec:cbr}
In this work, we introduce a novel geographic routing primitive called the \emph{Centroid}.  In physics, the centroid is defined as ``the center of mass of a geometric object of uniform density'' and our \emph{Centroid} intentionally evokes this idea in the context of geographic routing.  In looking at the location history of a mobile node, whether a circuit, linear path, or other arbitrary trace, we can envision a central point at which that trail would be balanced.  Unlike the physical centroid, which takes into account all the mass composing an object, we only concern ourselves with the points on the trace itself, which indeed may not for a closed shape at all.  This centroid then may be calculated as:

\begin{equation}
C_x(t_p)=\sum_{t=1}^{t_p}\frac{C_x(t-1)\times(t-1)}{t}+\frac{x_t}{t}
\end{equation}

Where $C$ is the Centroid, $x$ is the X, Y, or Z component of the node position, and $t_p$ is the present time increment.  Time increments are in terms of the chosen update interval.  In practice, we calculate the delta between the old Centroid and the new Centroid at each update as follows:

\begin{equation}
\Delta C_x(t_p)=\frac{x_{t_p}-C_x(t_p-1)}{t_p}
\end{equation}

This primitive is naturally resistant to noise introduced into the location history due to GPS reception errors, since they will be averaged out over time.  This is in contrast to positional routing primitives that rely only few/recent GPS readings to make routing decisions, and in some cases apply transformations such as trajectory calculation that amplify the effects of errors in those readings.

As with probabilistic and other routing protocols that predict a node's behavior based on past locations and encounters, the primary assumption with the Centroid is that a node's behavior will have repetitive qualities, so it is not suitable for one-shot type mobility patterns.  To go from primitive to protocol, then, there are many possible paths.  We explore a couple of these is this paper and leave others to future work.

\subsection{Goals}
We have a few goals for our protocols:
\begin{enumerate}
\item Resilience to noise/error found in positional data
\item Low power consumption
\item Simplicity; at this point we want to evaluate the Centroid primitive, not overshadow it with complex behaviors
\end{enumerate}

The first and third are relatively self-explanatory, but the second deserves some additional discussion.  While there are many factors that influence power-consumption, in this context we assume that avoiding unnecessary packet transmissions will have the greatest impact. 
We also observe that some DTN protocols are explicitly designed to maximize use of resources in order to improve the probability of delivery messages.  This is entirely reasonable given certain assumptions, such as vehicular networking where power is essentially unlimited with respect to communications.  Under other assumptions, such as personal devices and low-power sensors this approach is not optimal.  In our case, we are designing for the later case, which also corresponds to devices that commonly have cheap GPS receivers and compromised antennas.  Another consideration is that op-in users are less likely to forward packets if it noticeably drains the battery on their device.  For these reasons we want to explicitly conserve resources where possible.

\subsection{Centroid Router}
One mechanism that has shown significant promise in prior investigations is that employed by the Vector routing protocol, where the number of messages exchanged at an encounter is proportional to the orthogonality of the two nodes' trajectories.  Unfortunately, as described earlier, projecting a trajectory amplifies the effects of positional measurement errors.  We propose the Centroid-based analog of this, where the number of messages exchanged at an encounter is proportional to the cartesian distance between the two nodes' Centroids.  We call the implementation of this the Centroid Router.

At each encounter, the Centroid Router exchanges its current centroid, message list (list of all messages currently in the node's buffer), ACK list (list of all messages that have been acknowledged by the recipient), and current neighbor list.  ACK'd messages are deleted from each nodes' buffer first, and message exchange is begun with delivery of messages addressed to the nodes exchanging the messages.  The next set of messages exchanged are those addressed to currently connected neighbors of the two nodes.  None of these messages count against the message limit.  If this is all completed and the link is still established, the nodes proceed to exchange other messages subject to the message limit.  To find the limit, each node calculates the Centroid Distance, which is the distance between its own Centroid and the Centroid of the node it is exchanging messages with.  It then finds the ratio between this distance and the longest Centroid Distance it has previously calculated for all node encounters.  The resulting ratio is the fraction of its own message list that it is allowed to send to the connected neighbor.  This results in a linear relationship between Centroid Distance and fraction of messages exchanged.  More messages are exchanged with nodes that have more distant centroids, thus spreading messages as widely as possible with fewer forwarding events.

\subsection{CenterMass Router}
The CenterMass Router begins with the Centroid Router, and adds one additional mechanism, which is to forward messages \emph{in the direction of} their destination.  To achieve this, each node maintains a list of the Centroids for all nodes it has encountered, along with the address of each node and timestamp of the encounter.  At each encounter, Centroid lists are exchanged and merged into the nodes' own Centroid list.  In the case of a collision for a particular node address, the node keeps only the newer entry based on the timestamps.

When forwarding messages, the message limit is calculated as in the Centroid Router, however messages are only forwarded if the distance between the neighbor's Centroid and the destination's Centroid is smaller than the distance between the current host's Centroid and the destination's Centroid.  This mechanism minimizes the spread of messages in a direction away from the destination.

\section{Simulations \& Analysis}\label{sec:sa}

\begin{table}[t]
\centering
\caption{HELSINKI SCENARIO PARAMETERS}\label{tab:helsinki-params}
\begin{tabular}{| l | r |}\hline
\textbf{Parameter} & \textbf{Value}\\
\hline\hline
simulated duration & 12 hrs\\
\hline
warmup time & 1000 s\\
\hline
timestep resolution & 0.1 s\\
\hline
number of runs & 4\\
\hline
radio bandwidth & 10 Mb/s\\
\hline
transmit range & 10 m\\
\hline
buffer size & 5 MB\\
\hline
number of pedestrians & 80\\
\hline
pedestrian speed & 0.5--1.5 m/s\\
\hline
pedestrian pause time & 0--120 s\\
\hline
number of cars & 40\\
\hline
car speed & 2.7--13.9 m/s\\
\hline
car pause time & 0--120 s\\
\hline
number of trams & 6\\
\hline
tram speed & 7--10 m/s\\
\hline
tram pause time & 10--30 s\\
\hline
message rate & 1 / 25--35 s\\
\hline
message size & 0.5--1.0 MB\\
\hline
message TTL & 5 hrs \\
\hline
\end{tabular}
\end{table}

\begin{figure*}[t]
\centering
\subfigure[Delivery probability vs. radio bandwidth]{
\includegraphics[width=.31\linewidth]{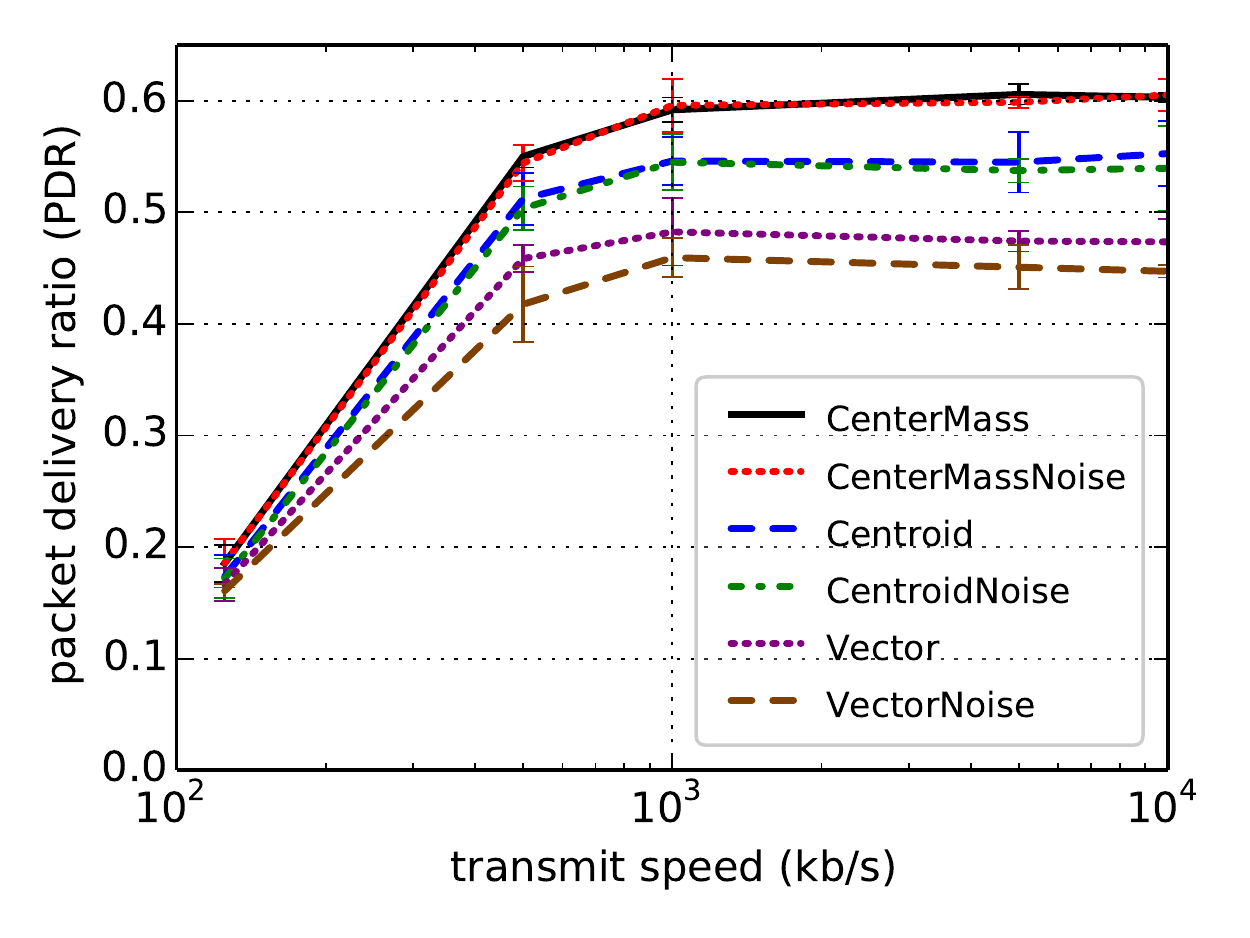}
\label{fig:helsinki_noise_dp_vs_ts_lc}}
\hfill
\subfigure[Average latency vs. radio bandwidth]{
\includegraphics[width=.31\linewidth]{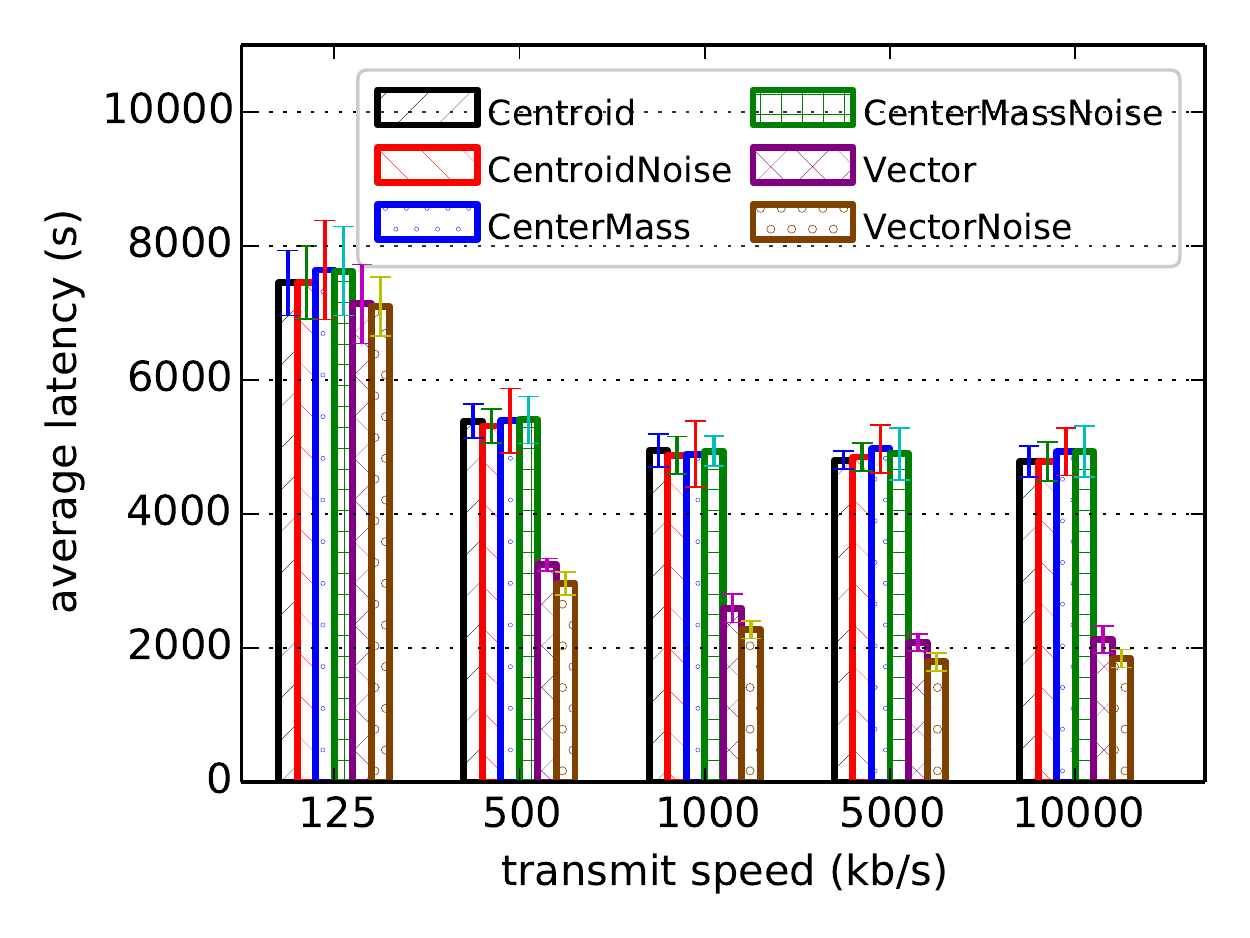}
\label{fig:helsinki_noise_la_vs_ts_lc}}
\hfill
\subfigure[Overhead ratio vs. buffer size]{
\includegraphics[width=.31\linewidth]{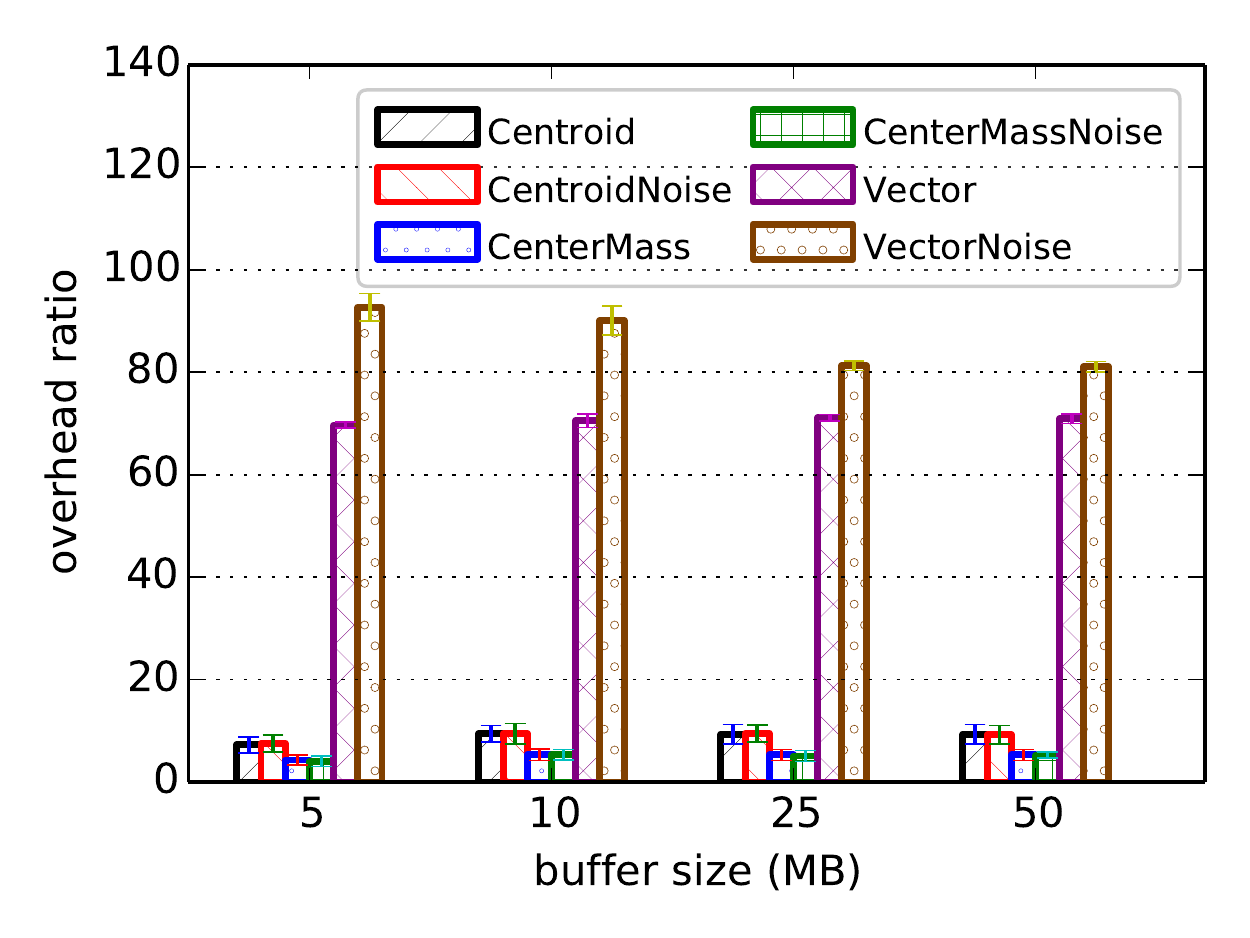}
\label{fig:helsinki_noise_overhead_vs_bs_lc}}
\hfill
\centering
\subfigure[Overhead ratio vs. radio bandwidth]{
\includegraphics[width=.31\linewidth]{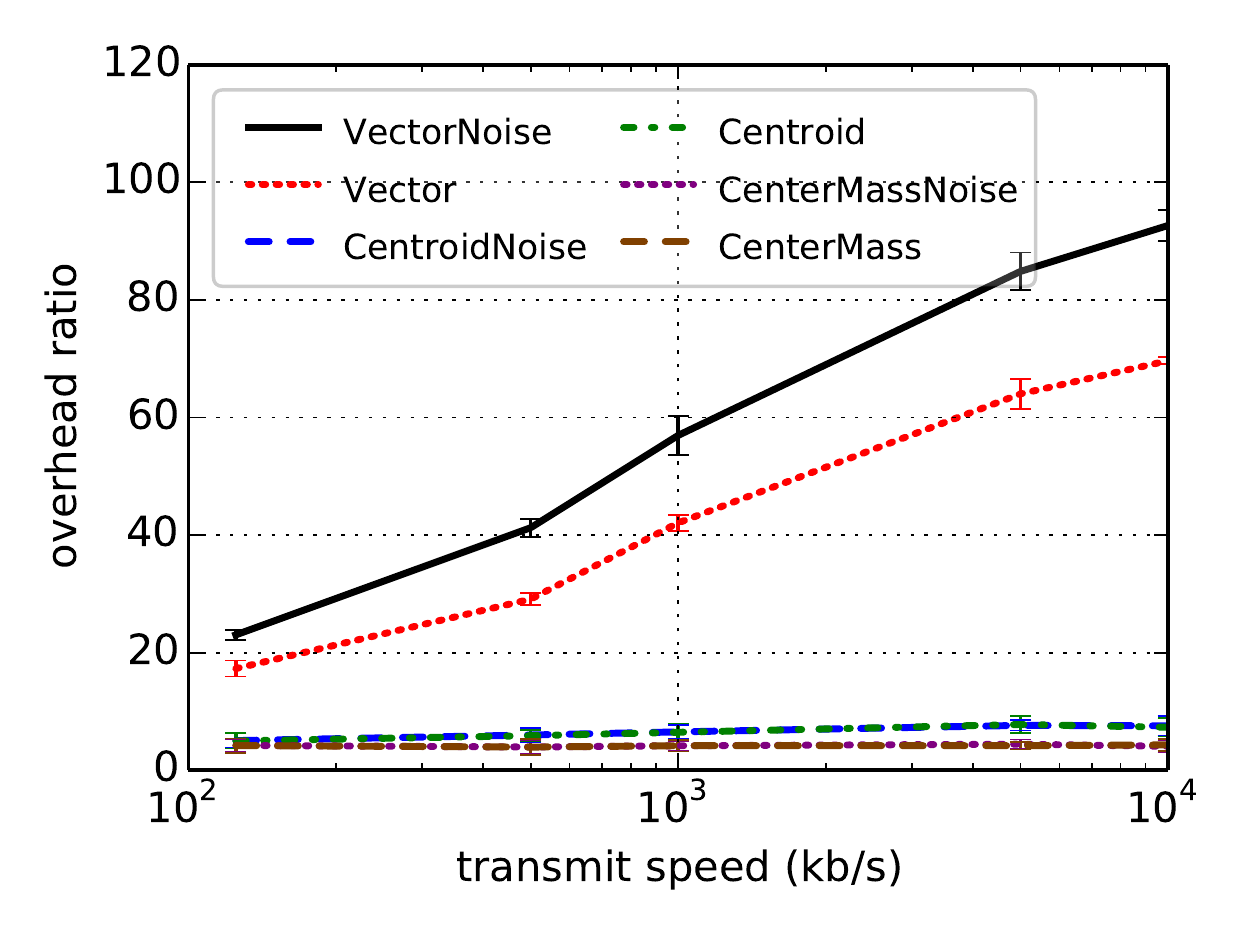}
\label{fig:helsinki_noise_overhead_vs_ts_lc}}
\hfill
\subfigure[Delivery probability vs. radio bandwidth]{
\includegraphics[width=.31\linewidth]{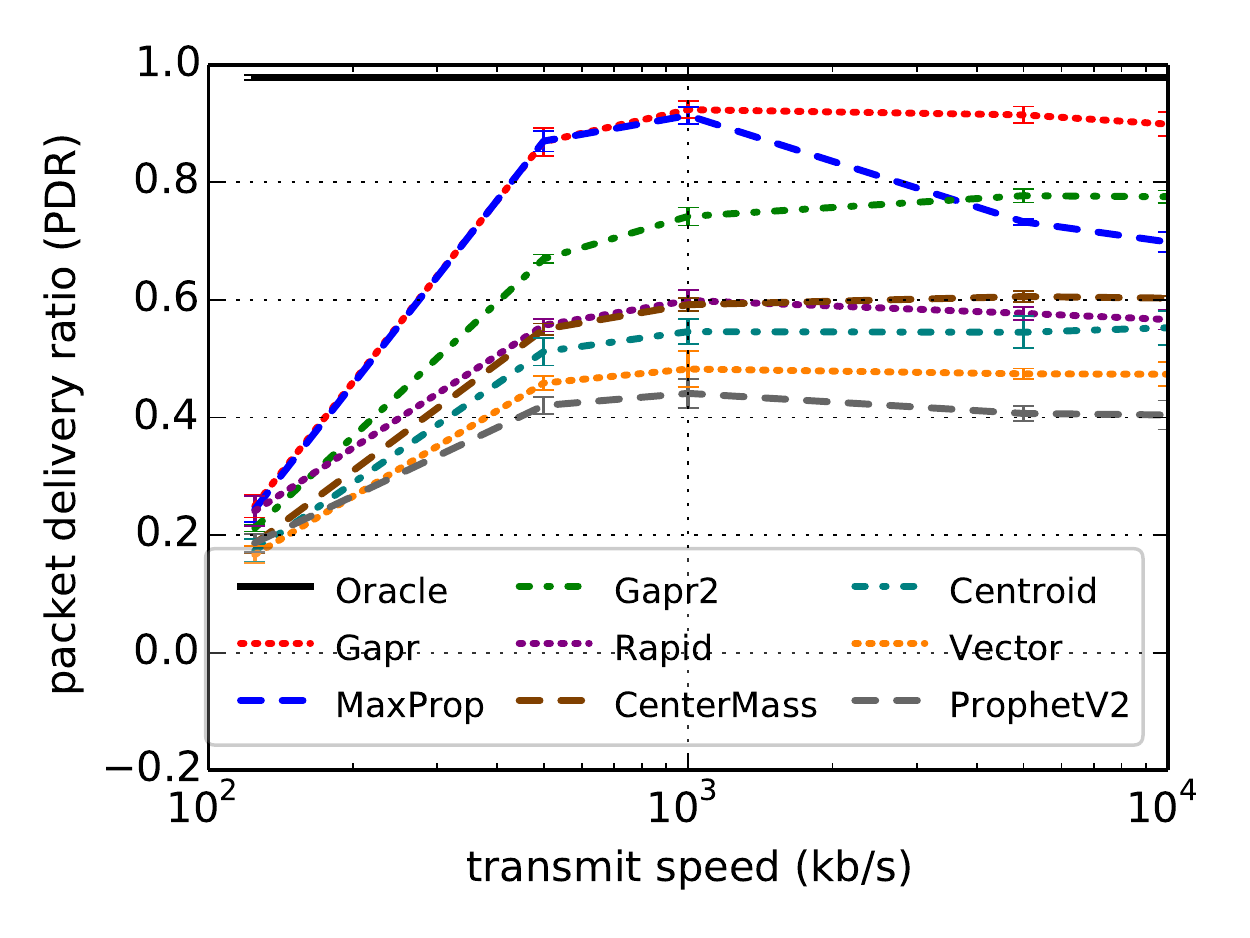}
\label{fig:helsinki_dp_vs_ts_lc}}
\hfill
\subfigure[Delivery probability vs. buffer size]{
\includegraphics[width=.31\linewidth]{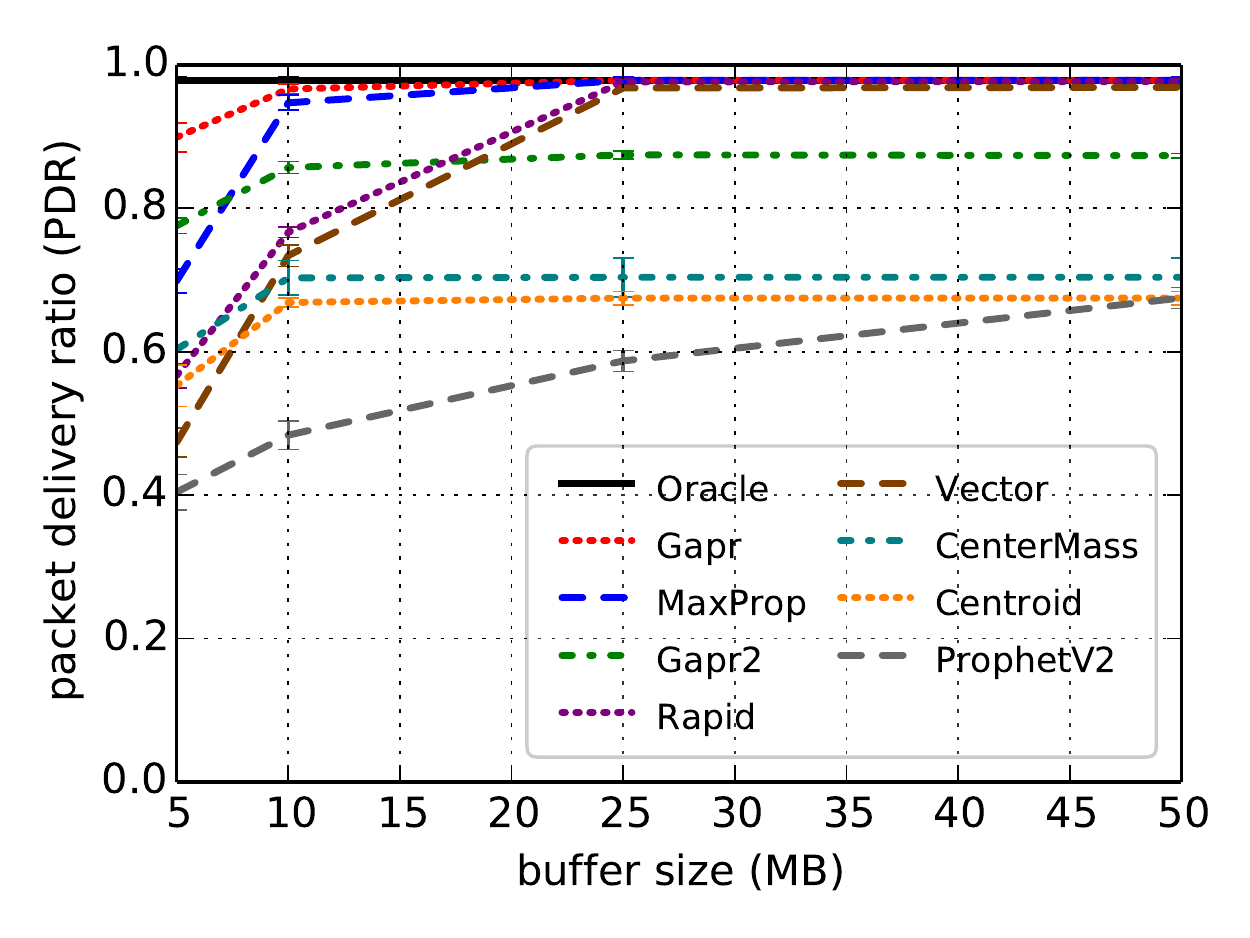}
\label{fig:helsinki_dp_vs_bs_lc}}
\caption{Performance effects of GPS errors}
\end{figure*}

\begin{figure*}[t]
\centering
\subfigure[Average latency vs. buffer size]{
\includegraphics[width=.31\linewidth]{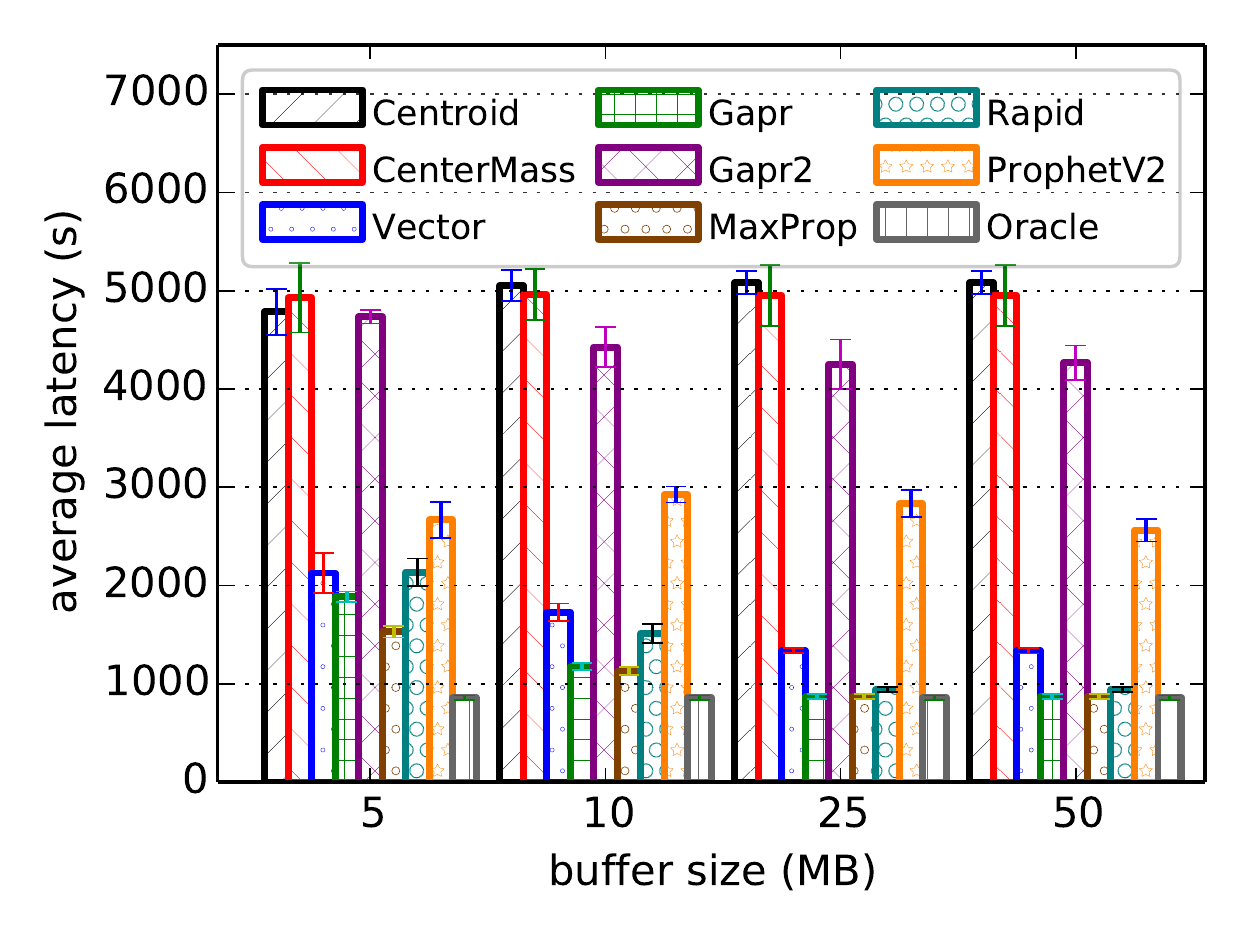}
\label{fig:helsinki_la_vs_bs_lc}}
\hfill
\subfigure[Average latency vs. radio bandwidth]{
\includegraphics[width=.31\linewidth]{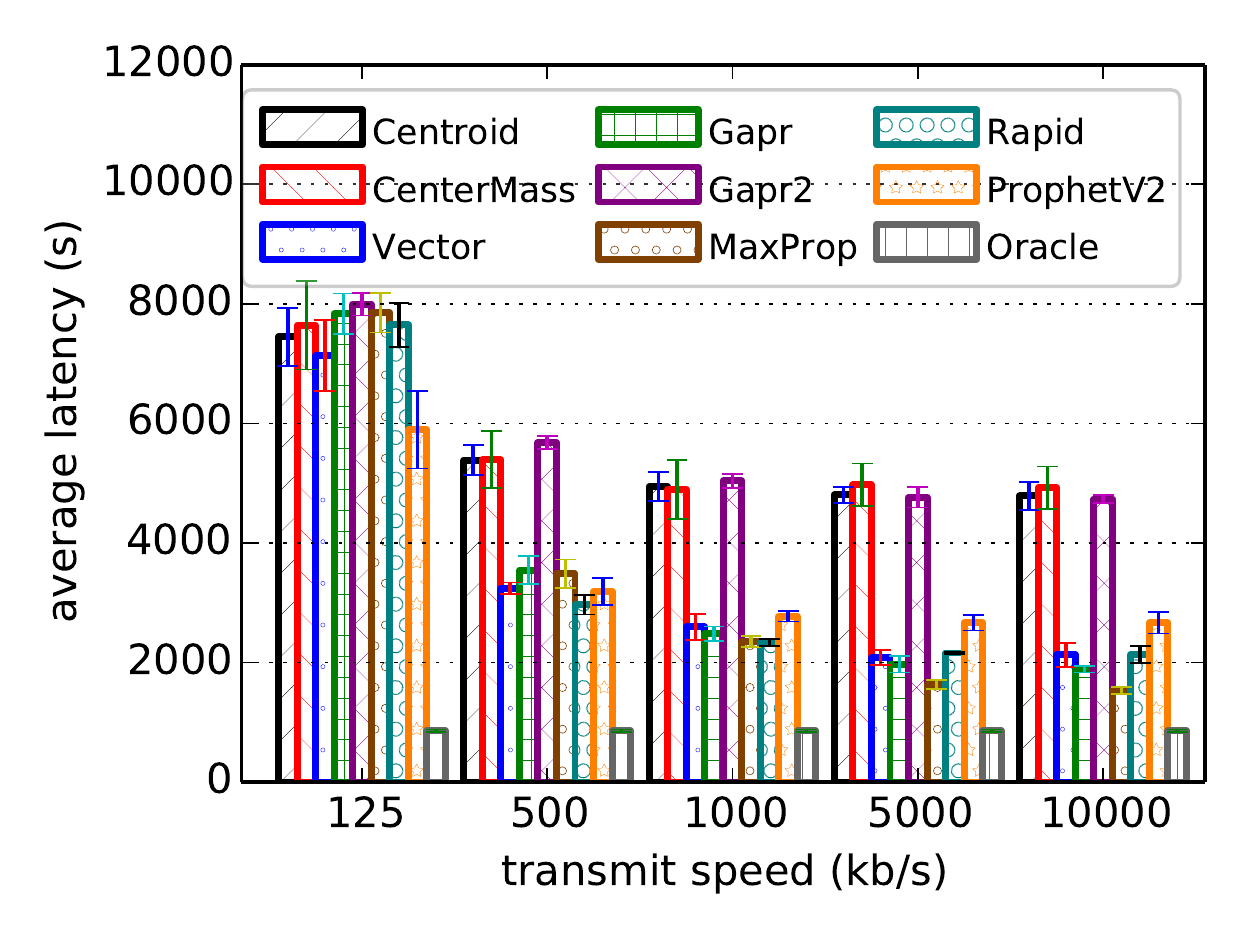}
\label{fig:helsinki_la_vs_ts_lc}}
\hfill
\subfigure[Overhead ratio vs. buffer size]{
\includegraphics[width=.31\linewidth]{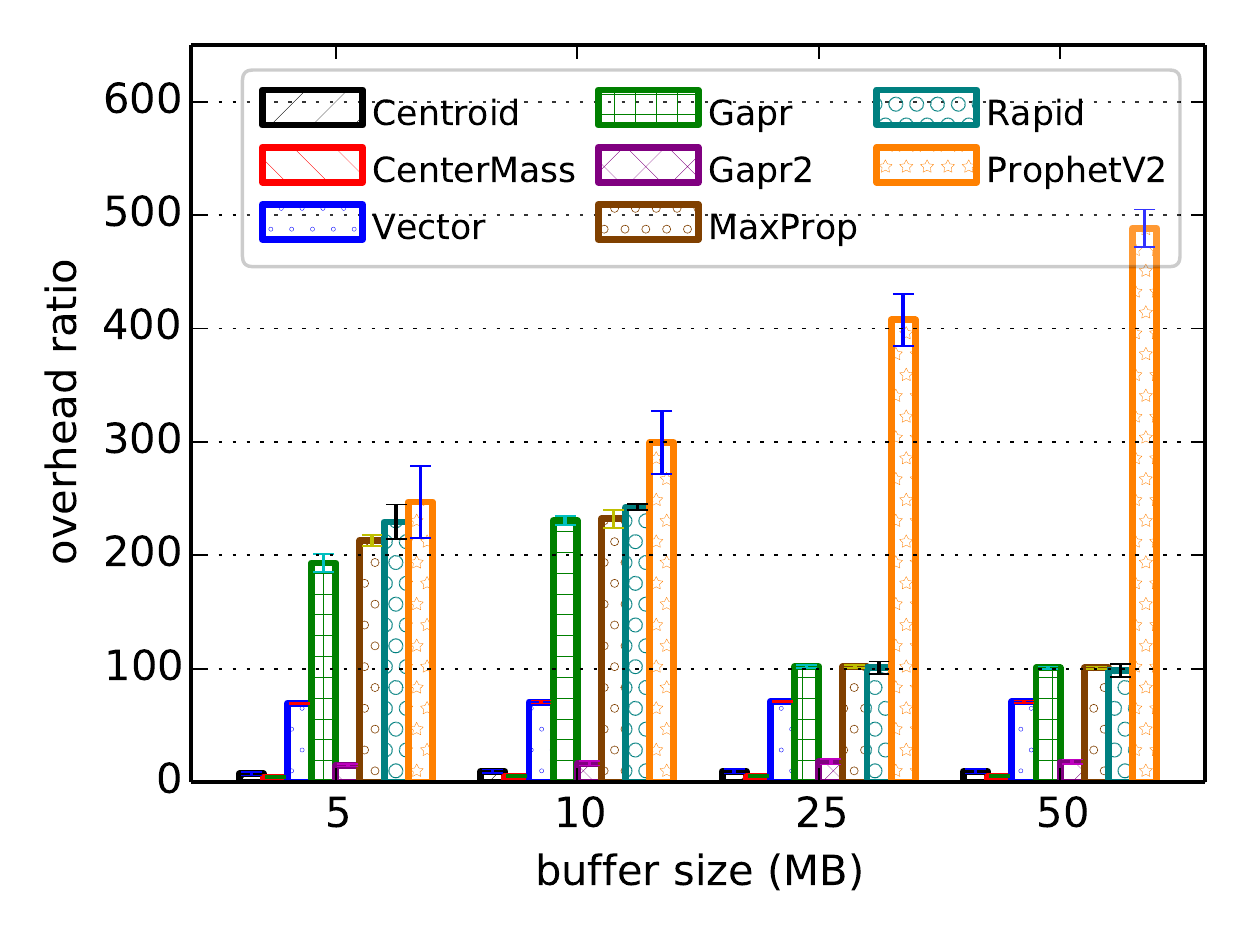}
\label{fig:helsinki_overhead_vs_bs_lc}}
\hfill
\subfigure[Overhead ratio vs. radio bandwidth]{
\includegraphics[width=.31\linewidth]{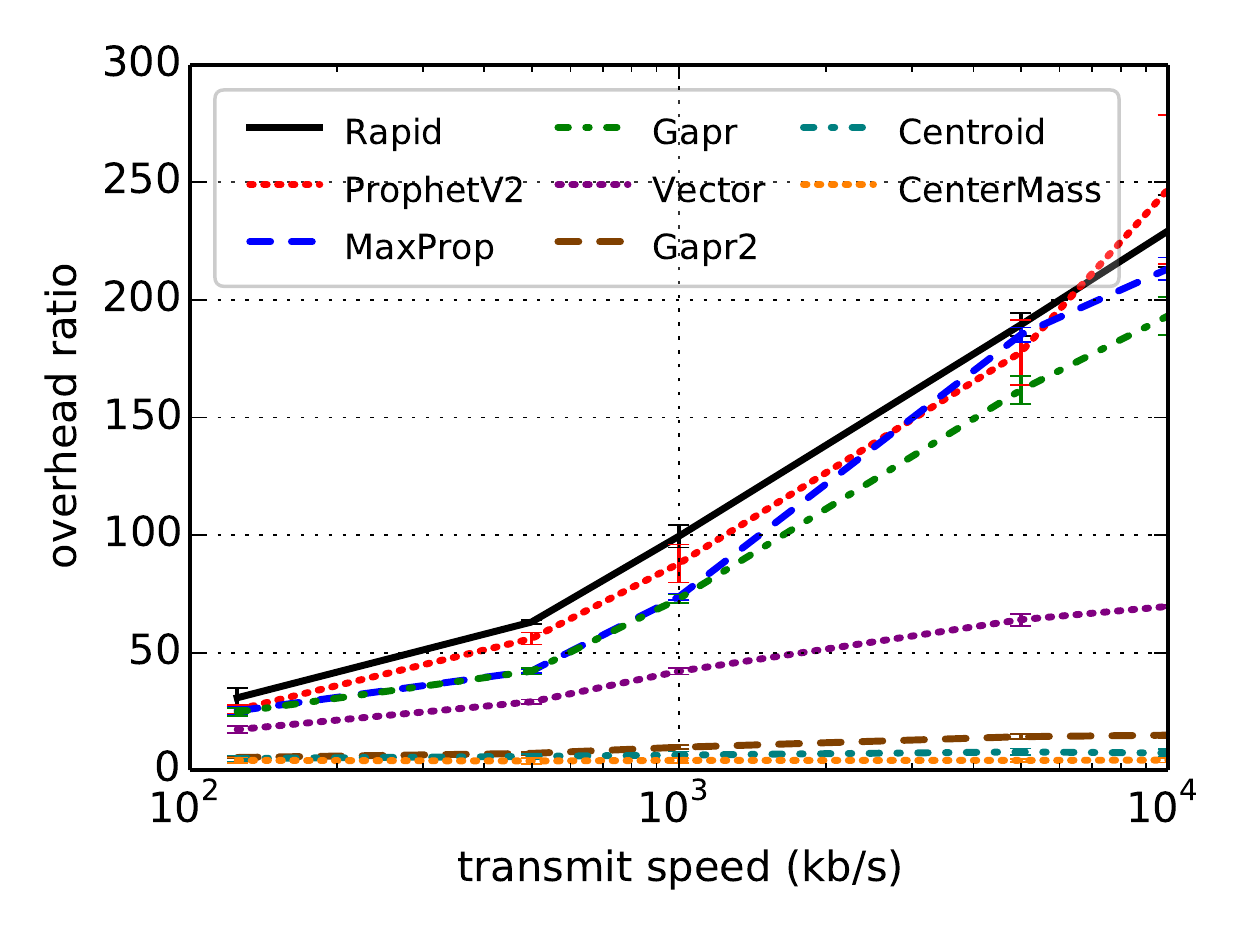}
\label{fig:helsinki_overhead_vs_ts_lc}}
\hfill
\subfigure[Protocol efficacy vs. buffer size]{
\includegraphics[width=.31\linewidth]{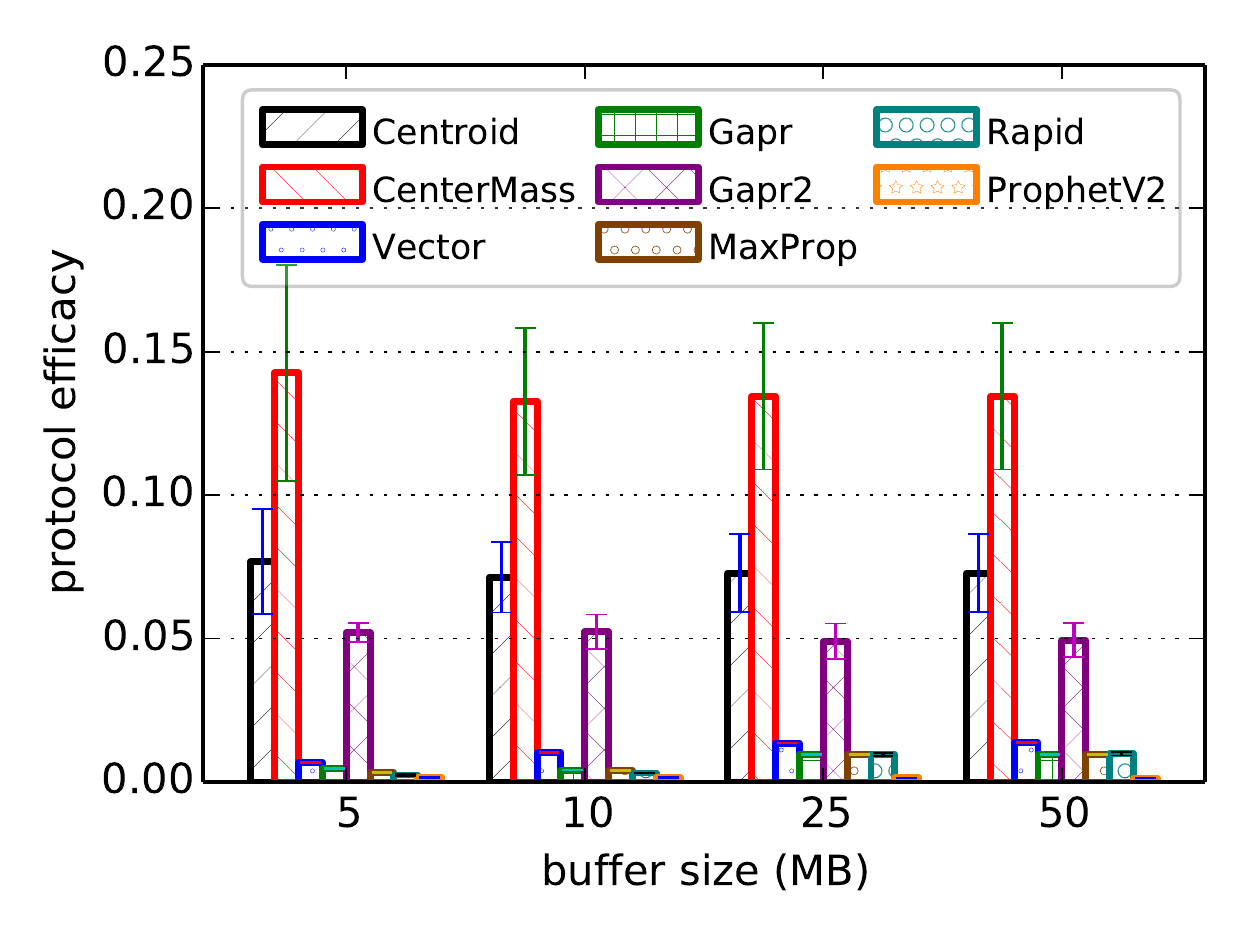}
\label{fig:helsinki_efficacy_vs_bs_lc}}
\hfill
\subfigure[Protocol efficacy vs. radio bandwidth]{
\includegraphics[width=.31\linewidth]{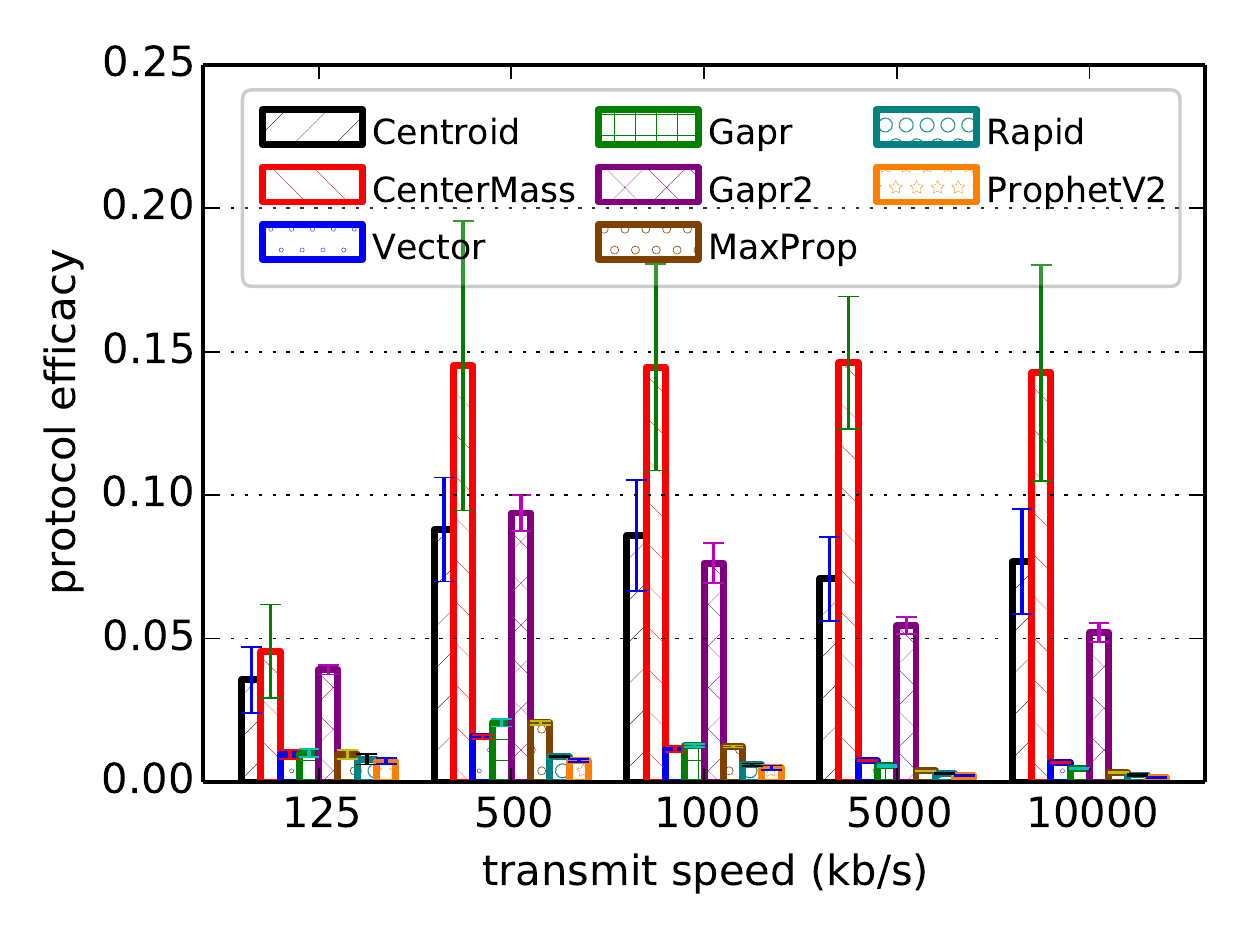}
\label{fig:helsinki_efficacy_vs_ts_lc}}
\caption{Helsinki scenario performance metrics}
\end{figure*}

We perform our analysis using The ONE Simulator~\cite{keranen:2009:TOS}, as it is specifically suited to DTN routing analysis.  In previous work, we have used a number of mobility scenarios with the ONE, however in this case evaluating the protocols on multiple scenarios did not yield any additional insights, so for clarity we present a single evaluation scenario in this work.  We choose the Helsinki map-based model, which has become well-known in DTN routing literature due to its inclusion as the default mobility model for the ONE simulator.  The model includes both vehicles and pedestrians that participate as nodes in the routing protocol.  We have made some minor changes to the default parameter values, which are shown in Table~\ref{tab:helsinki-params}.  To generate traffic, one random node sends a message to a random destination every 25--35 seconds.  Our simulation study consists of three components:
\begin{enumerate}
\item Evaluating the effect of positional measurement errors
\item Comparing our protocols' performance to more sophisticated probabilistic protocols that rely on encounter history to predict path costs
\item Discussing design tradeoffs and attempting to quantify their effect
\end{enumerate}

Each data point in the plots that follow represents the average of 4 simulation runs with varying random seeds, and the error bars on all the plots in this paper represent 95\% confidence intervals.

\subsection{Positional Sample Error}
As discussed earlier, we are concerned with the effect of errors in the positional (e.g. GPS) sample data provided to the routing protocol.  To evaluate this, we create alternate versions of both of our protocols (Centroid and CenterMass) as well as the Vector routing protocol, which add noise to the position provided by the simulator before using it in calculating their respective routing primitives.  This noise is random and uniform, in the range $\pm20$ m.  Note that this roughly the \emph{advertised} error for civilian GPS, and is far from a worst-case scenario that could be $\pm200$ m with strong correlation between samples.

Figure~\ref{fig:helsinki_noise_dp_vs_ts_lc} shows how this measurement error affects the Packet Delivery Ratio (PDR) of the three protocols.  All the protocols are bunched together at low (125 Kb/s) radio data rates.  With higher transmission rates (500 Kb/s -- 10 Mb/s) the protocols become distinguishable.  The Vector protocol is most significantly affected, with the positional errors noticeably reducing the packet delivery ratio.  That being said, the reduction is only about 10\% at its worst.  We believe that since the Vector protocol only relies on the trajectory to enable efficient spreading of messages, it may be less affected than a protocol that uses trajectory in a more specific manner (e.g. identifying trajectory in the direction of the message destination).  Unfortunately, we do not have such a protocol implemented in the ONE at this time to test our hypothesis.  Both of the Centroid-based protocols show negligible effects from the noise, as expected.  While the traces with noise trend lower than those without, they are within the 95\% confidence intervals of each other at almost every data point.  Not only is the Centroid routing protocol less affected by noise, it outperforms the Vector protocol by about 20\% in the presence of positional errors.  The CenterMass protocol achieves an additional 10\% performance improvement over Centroid.

We next examine the effect on latency, shown in Figure~\ref{fig:helsinki_noise_la_vs_ts_lc}.  The effects are very small across the board, but surprisingly Vector's latency \emph{improves} in the presence of errors.  Vector also outperforms the Centroid-based protocols and higher transmission rates.  Lastly, we look at the effect on overhead.  The overhead ratio reported the the ONE simulator is: $\frac{(\textrm{forwarded messages} - \textrm{delivered messages})}{\textrm{delivered messages}}$.  From Figure~\ref{fig:helsinki_noise_overhead_vs_bs_lc} we can now explain the reduced latency achieved by the Vector protocol, since there are literally $10\times$ more copies of every packet forwarded in the Vector routing simulations that there are in the Centroid routing simulations, and the positional errors make the Vector overhead approximately 30\% worse.  Increasing the buffer size reduces the impact of positional errors on Vector's overhead, and has almost no effect on Centroid or CenterMass.  We do note that in addition to the improved delivery probability of CenterMass over Centroid, CenterMass has significantly lower overhead than Centroid.  For a view of the effects of increasing transmission speed we look to Figure~\ref{fig:helsinki_noise_overhead_vs_ts_lc}.  Here, we see that not only does the effect of positional errors on Vector increase as more bandwidth is made available, but the absolute overhead appears to run-away, quadrupling between 125 Kb/s and 10 Mb/s.  Centroid also has increased overhead as the bandwidth increases, but only slightly, and there appears to be almost no effect on the overhead of CenterMass.

From these plots, we see that the negative impact of positional errors on some protocols is real, and that these two Centroid-based routing protocols have a significant advantage in terms of overhead, relative to Vector, a protocol of comparable complexity and message delivery performance.

\subsection{Probabilistic-Predictive Routing Comparisons}
Having satisfied the question of whether the Centroid primitive minimizes the effects of positional error, we can then proceed to comparisons with a larger selection of routing protocols in under ideal (no GPS error) circumstances confident that our protocols will not degrade with real positional data and offering a number of other protocols a best-case scenario of no errors (and indeed several of them do not rely on positional data anyway).  Here, we are looking to see how our relatively simple protocols compare to much more sophisticated protocols that use encounter history to predict probability of future delivery, including a couple from our own prior work (Geolocation Assisted Routing Protocol (GAPR) and GAPR2)~\cite{rohrer:2016:GAR}.  For this comparison we have also created an Oracle router that demonstrates the best possible delivery ratio given the scenario contact graph.  Figure~\ref{fig:helsinki_dp_vs_ts_lc} shows these results, with respect to radio transmission rate.  One feature of particular note, is that while most protocols reach a maximum delivery probability and then plateau, MaxProp~\cite{burgess:2006:MRF}, though it ties for the highest overall delivery ratio, decreases significantly (20\%) with \emph{higher} transmission rates.  We do not have an explanation for this, but have observed it in the past.  Aside from this anomaly, MaxProp and our own GAPR and GAPR2 protocols all out perform CenterMass and Centroid, by as much as 30\%.  CenterMass is roughly equivalent to Rapid~\cite{balasubramanian:2007:DRA}, both of which are about 10\% improved over Centroid, with Vector and PRoPHETV2 10 and 20\% worse than Centroid, respectively.  Taken by itself, this shows that the sophistication of probabilistic protocols is not without merit.  Figure~\ref{fig:helsinki_dp_vs_bs_lc} paints a similar picture, but also shows that with increase available buffer space some of the lesser performing protocols in the previous plot show great improvement, notably Rapid and Vector, both of which approach the performance of the Oracle with sufficient buffer availability.  Unfortunately, neither of the Centroid-based protocols fall into this category, and their only redeeming quality in this plot is that they continue to outperform PRoPHETv2~\cite{grasic:2011:TEO}.  Fortunately, that is not the end of the analysis!

Pressing on we examine latency in Figures~\ref{fig:helsinki_la_vs_bs_lc} and~\ref{fig:helsinki_la_vs_ts_lc}.  Here, we see that CenterMass and Centroid are at the higher-end of the latency spectrum, along with GAPR2, while our older protocol GAPR can acheive approximately the same latency as the Oracle, given sufficient buffer space.  We note that MaxProp, Vector and Rapid all perform well with respect to latency.  We also note again the trend of improvement with increased buffer sizes, which leads us to the topic of overhead.

Examining Figure~\ref{fig:helsinki_overhead_vs_bs_lc} we see that the Centroid-based protocols have overhead ratios of 10 or below, and are invariant to buffer size.  With the exception of GAPR2, the other protocols are about 1 order of magnitude higher, with PRoPHETv2's overhead increasing dramatically at higher buffer sizes.  In Figure~\ref{fig:helsinki_overhead_vs_ts_lc} we see that Vector's runaway overhead was relatively minor compared to most of the other protocols whose overhead appears to increase exponentially in some cases as transmission rate increases.  As before, the overhead of Centroid and CenterMass are small and relatively unaffected by the radio bandwidth.

\subsection{Protocol Efficacy}
Given that the message delivery performance of our Centroid-based protocols was outclassed by several of the probabilistic protocols, but they were the clear leader in terms of overhead, we would like to be able to show the combined effect using a single metric that captures the goals we presented earlier on in the paper.  We would use efficiency, but it is roughly the inverse of overhead, and does not fully serve our purposes.  So, we introduce Protocol Efficacy, which is simply $\frac{(\textrm{delivery ratio})}{\textrm{overhead}}$.  In the ideal case where delivery ratio is 1 and overhead is 1, efficacy would also be 1, and any time delivery ratio is 0, efficacy is 0.  This metric then captures both the goal of high delivery ratio and low overhead.  Figures~\ref{fig:helsinki_efficacy_vs_bs_lc} and~\ref{fig:helsinki_efficacy_vs_ts_lc} show this metric for each of the protocols discussed so far.  We see that the CenterMass protocol outperforms the rest of the field by a significant margin, Centroid and GAPR2 are comparable to one another, and the rest (Vector, GAPR, MaxProp, Rapid, and PRoPHETv2) fall behind by and order of magnitude or more.  

\subsection{Observations}
The simulation of the Rapid routing protocol runs about two orders-of-magnitude slower than any of the other protocols tested.  Since the total number of messages forwarded in the Rapid simulations is not particularly high, we can only assume that the computations required are of significant complexity.  This may be of concern when trying to conserver power on low cost/performance devices.

We again note that some routing protocols are designed to continue transmitting as long as connection stays up, even in the face of diminishing marginal performance gains, based on the assumption that power (and therefore transmissions) are free.  This design choice results in the runaway overhead results seen above, but may not be of concern for certain environments.  Our design is different in that it explicitly stops transmitting to conserve resources when those transmissions are unlikely to result in message delivery.  This then reflects the difference is design philosophies between explicitly conserving resources (power) and maximizing resource use (bandwidth/buffer space).

\section{Conclusion}\label{sec:c}
We have demonstrated the negative effect that positional errors can have on routing protocols in a vehicular network that rely on geolocation inputs, depending on how that input is used.  We have also demonstrated Centroid Routing and CenterMass Routing, both of which are immune to random error in positional data inputs.  We have shown how these protocols out-perform existing state-of-the-art probabilistic routing protocols both in terms of traditional metrics and using our new Protocol Efficacy metric.  These new protocols show a dramatic improvement over existing protocols when normalized against the overhead they induce in the network.

We envision many possible applications of the Centroid primitive.  One of the simplest is as a direct substitute for position in position-based routing.  Another is as an enhancement to our own GAPR2 routing protocol, to replace the raw position data currently employed.  The CenterMass routing protocol also has potential for refinement, and we think that by combining geolocation and encounter history data in some meaningful way we can approach the delivery probability of probabilistic protocols while retaining the Efficacy of the Centroid-based protocols shown here.  We expect that the Efficacy metric will be particularly significant in evaluating high-volume, loss tolerant routing environments such as drone-swarms~\cite{pospischil:2017:MRO}.  In the future we intend to continue this work using a simulator with higher-fidelity network, MAC, and physical layer models, such as ns-3, in which we have begun implementing DTN routing protocols~\cite{rohrer:2018:IOE}.

\section*{Acknowledgment}
This work was funded in part by the US Marine Corps and US Navy. Views and conclusions are those of the authors and should not be interpreted as representing the official policies or position of the U.S. government.



\bibliographystyle{IEEEtran}
\bibliography{vehicular.bib,../bib/master,../bib/rfc,../bib/jprohrer,../bib/theses}

\begin{thebibliography}{10}
\providecommand{\url}[1]{#1}
\csname url@samestyle\endcsname
\providecommand{\newblock}{\relax}
\providecommand{\bibinfo}[2]{#2}
\providecommand{\BIBentrySTDinterwordspacing}{\spaceskip=0pt\relax}
\providecommand{\BIBentryALTinterwordstretchfactor}{4}
\providecommand{\BIBentryALTinterwordspacing}{\spaceskip=\fontdimen2\font plus
\BIBentryALTinterwordstretchfactor\fontdimen3\font minus
  \fontdimen4\font\relax}
\providecommand{\BIBforeignlanguage}[2]{{%
\expandafter\ifx\csname l@#1\endcsname\relax
\typeout{** WARNING: IEEEtran.bst: No hyphenation pattern has been}%
\typeout{** loaded for the language `#1'. Using the pattern for}%
\typeout{** the default language instead.}%
\else
\language=\csname l@#1\endcsname
\fi
#2}}
\providecommand{\BIBdecl}{\relax}
\BIBdecl

\bibitem{rohrer:2008:CLA}
\BIBentryALTinterwordspacing
J.~P. Rohrer, A.~Jabbar, E.~Perrins, and J.~P.~G. Sterbenz, ``Cross-layer
  architectural framework for highly-mobile multihop airborne telemetry
  networks,'' in Proceedings of the {IEEE} Military Communications Conference
  ({MILCOM}), San Diego, CA, USA, November 2008, pp. 1--9.
\BIBentrySTDinterwordspacing

\bibitem{benmoshe:2011:IAO}
B.~Ben-Moshe, E.~Elkin, H.~Levi, and A.~Weissman, ``Improving accuracy of
  {GNSS} devices in urban canyons,'' in Proceedings of the 23rd Canadian
  Conference on Computational Geometry ({CCCG}), August 10--12 2011, pp.
  399--404.

\bibitem{hang:2008:VRF}
H.~Kang and D.~Kim, ``Vector routing for delay tolerant networks,'' in
  Proceedings of the {IEEE} 68th Vehicular Technology Conference, Sept 2008,
  pp. 1--5.

\bibitem{liao:2001:GAF}
W.-H. Liao, J.-P. Sheu, and Y.-C. Tseng, ``{GRID}: A fully location-aware
  routing protocol for mobile ad hoc networks,'' Telecommunication Systems,
  vol.~18, no. 1-3, 2001, pp. 37--60.

\bibitem{fuente:2007:APC}
M.~G. de~la Fuente and H.~Ladiod, ``A performance comparison of position-based
  routing approaches for mobile ad hoc networks,'' Vehicular Technology
  Conference ({VTC}), October 2007, pp. 1--5.

\bibitem{galluccio:2007:AMR}
L.~Galluccio, A.~Leonardi, G.~Morabito, and S.~Palazzo, ``A {MAC}/routing
  cross-layer approach to geographic forwarding in wireless sensor networks,''
  Ad Hoc Netw., vol.~5, no.~6, 2007, pp. 872--884.

\bibitem{mauve:2001:ASO}
M.~Mauve, A.~Widmer, and H.~Hartenstein, ``A survey on position-based routing
  in mobile ad hoc networks,'' {IEEE} Network, vol.~15, no.~6, 2001, pp.
  30--39.

\bibitem{yuksel:2006:AIF}
M.~Yuksel, R.~Pradhan, and S.~Kalyanaraman, ``{An implementation framework for
  trajectory-based routing in ad hoc networks},'' Ad Hoc Networks, vol.~4,
  no.~1, 2006, pp. 125--137.

\bibitem{iordanakis:2006:ARP}
M.~Iordanakis et~al., ``Ad-hoc routing protocol for aeronautical mobile ad-hoc
  networks,'' in Proceedings of the Fifth International Symposium on
  Communication Systems, Networks and Digital Signal Processing ({CSNDSP}),
  2006, pp. 1--5.

\bibitem{jabbar:2009:AAG}
A.~Jabbar and J.~P.~G. Sterbenz, ``{AeroRP}: A geolocation assisted
  aeronautical routing protocol for highly dynamic telemetry environments,'' in
  Proceedings of the International Telemetering Conference ({ITC}), Las Vegas,
  NV, October 2009, pp. 1--10.

\bibitem{rohrer:2011:HDC}
\BIBentryALTinterwordspacing
J.~P. Rohrer, A.~Jabbar, E.~K. \c{C}etinkaya, E.~Perrins, and J.~P. Sterbenz,
  ``Highly-dynamic cross-layered aeronautical network architecture,'' {IEEE}
  Transactions on Aerospace and Electronic Systems ({TAES}), vol.~47, no.~4,
  October 2011, pp. 2742--2765.
\BIBentrySTDinterwordspacing

\bibitem{sterbenz:2018:DTA}
\BIBentryALTinterwordspacing
J.~P.~G. Sterbenz et~al., ``Disruption-tolerant airborne networks and
  protocols,'' in {UAV} Networks and Communications, 1st~ed., K.~Namuduri,
  S.~Chaumette, J.~H. Kim, and J.~P.~G. Sterbenz, Eds.\hskip 1em plus 0.5em
  minus 0.4em\relax Cambridge University Press, January 2018, ch.~4, pp.
  58--95.
\BIBentrySTDinterwordspacing

\bibitem{ko:2000:LAR}
Y.-B. Ko and N.~H. Vaidya, ``Location-aided routing {(LAR)} in mobile ad hoc
  networks,'' {Journal of Wireless Networks}, vol.~6, no.~4, Jul. 2000, pp.
  307--321.

\bibitem{sanchez:2009:BGR}
J.~Sanchez, P.~Ruiz, and R.~Marin-Perez, ``Beacon-less geographic routing made
  practical: Challenges, design guidelines, and protocols,'' {IEEE}
  Communications Magazine, vol.~47, no.~8, August 2009, pp. 85--91.

\bibitem{blum:2003:IAS}
B.~Blum, T.~He, S.~Son, and J.~Stankovic, ``{IGF}: A state-free robust
  communication protocol for wireless sensor networks,'' Department of Computer
  Science, University of Virginia, USA, Technical Report CS-2003-11, 2003.

\bibitem{sanchez:2007:BBO}
J.~Sanchez, R.~Marin-Perez, and P.~Ruiz, ``{BOSS}: Beacon-less on demand
  strategy for geographic routing in wireless sensor networks,'' in {IEEE}
  Internatonal Conference on Mobile Adhoc and Sensor Systems ({MASS}), October
  2007, pp. 1--10.

\bibitem{heissenbuttel:2004:BBR}
M.~Heissenb{\"u}ttel, T.~Braun, T.~Bernoulli, and M.~W{\"a}lchli, ``{BLR}:
  Beacon-less routing algorithm for mobile ad hoc networks,'' Computer
  Communications, vol.~27, no.~11, 2004, pp. 1076--1086.

\bibitem{rohrer:2017:EOG}
\BIBentryALTinterwordspacing
J.~P. Rohrer, ``Effects of {GPS} error on geographic routing,'' in Proceedings
  of the 26th International Conference on Computer Communications and Networks
  ({ICCCN}).\hskip 1em plus 0.5em minus 0.4em\relax Vancouver, Canada: IEEE,
  August 2017, pp. 1--2.
\BIBentrySTDinterwordspacing

\bibitem{keranen:2009:TOS}
A.~Ker\"{a}nen, J.~Ott, and T.~K\"{a}rkk\"{a}inen, ``The {ONE} simulator for
  {DTN} protocol evaluation,'' in Proceedings of the 2nd International
  Conference on Simulation Tools and Techniques ({SIMUTools}).\hskip 1em plus
  0.5em minus 0.4em\relax New York, NY, USA: ICST, 2009, pp. 55:1--55:10.

\bibitem{rohrer:2016:GAR}
\BIBentryALTinterwordspacing
J.~P. Rohrer and K.~M. Killeen, ``Geolocation assisted routing protocols for
  vehicular networks,'' in Proceedings of the 5th {IEEE} International
  Conference on Connected Vehicles ({ICCVE}), Seattle, WA, September 2016, pp.
  1--6.
\BIBentrySTDinterwordspacing

\bibitem{burgess:2006:MRF}
J.~Burgess, B.~Gallagher, D.~Jensen, and B.~N. Levine, ``Maxprop: Routing for
  vehicle-based disruption-tolerant networks,'' in Proceedings of the 25th
  {IEEE} International Conference on Computer Communications {(INFOCOM)}, April
  2006, pp. 1--11.

\bibitem{balasubramanian:2007:DRA}
A.~Balasubramanian, B.~Levine, and A.~Venkataramani, ``{DTN} routing as a
  resource allocation problem,'' in Proceedings of the {ACM} conference on
  Applications, technologies, architectures, and protocols for computer
  communications ({SIGCOMM}), vol.~37, Oct. 2007, pp. 373--384.

\bibitem{grasic:2011:TEO}
S.~Grasic, E.~Davies, A.~Lindgren, and A.~Doria, ``The evolution of a {DTN}
  routing protocol -- {PRoPHETv2},'' in Proceedings of the 6th {ACM} Workshop
  on Challenged Networks (CHANTS).\hskip 1em plus 0.5em minus 0.4em\relax ACM,
  Sep. 2011, pp. 27--30.

\bibitem{pospischil:2017:MRO}
\BIBentryALTinterwordspacing
A.~Pospischil and J.~P. Rohrer, ``Multihop routing of telemetry data in drone
  swarms,'' in Proceedings of the International Telemetering Conference
  ({ITC}), Las Vegas, NV, October 2017, pp. 1--10.
\BIBentrySTDinterwordspacing

\end{thebibliography}
%
%
%

\end{document}